\documentclass[]{imsart}

\usepackage[latin1]{inputenc}
\usepackage[cyr]{aeguill}

\usepackage{amsthm,amsmath}
\RequirePackage[colorlinks,citecolor=blue,urlcolor=blue]{hyperref}

\usepackage{hypernat}

\usepackage{amsfonts}%
\usepackage{amssymb}%
\usepackage{graphicx}

\startlocaldefs

\theoremstyle{plain}

\newtheorem{definition}{Definition}

\newtheorem{remark}{Remark}

\numberwithin{equation}{section}
\endlocaldefs

\begin{document}
\begin{frontmatter}

\title{A definable number which cannot be approximated algorithmically }
\runtitle{A definable number which cannot be approximated algorithmically}


\author{\fnms{Nicolas} \snm{Brener} \ead[label=e1]{brener@isir.upmc.fr}}
\address{Institut des Systèmes Intelligents et de Robotique\\ Université Pierre et Marie Curie\\ 4 Place Jussieu, 75252 Paris Cedex 05\\
\printead{e1}}



\runauthor{N. Brener}

\begin{abstract}
The Turing machine (TM) and the Church thesis have formalized the concept of computable number, this allowed to display non-computable numbers. This paper defines the concept of number ``approachable'' by a TM and shows that some (if not all) known non-computable numbers are approachable by TMs. Then an example of a number not approachable by a TM is given.
\end{abstract}





\end{frontmatter}

\section{Introduction}

The Turing machine (TM) introduced in \cite{turing} and the thesis of Church-Turing formalized the concept of algorithm and computable number. As a consequence it became possible to display non-computable real numbers. However, even when a real number is not computable it may be the limit of a computable function, then we say that it is ``approachable''. Some well known non-computable numbers are approachable by a TM. But to my knowledge there is no example of a number for which it was established that it is not approachable by a TM. The purpose of this paper is to produce a real number that is not the limit of a computable function.

The concept of computable number and the principle to derive a non-computable number as formerly proposed by Turing\cite{turing} are recalled in section §\ref{sec:calcul}. 
The section §\ref{sec:app} defines the concept of ``approachable'' real number. Relying on this definition, section §\ref{sec:appnoncalc} displays a non-computable number which is approachable. Finally, section §\ref{sec:nonapp} gives an example of a non-approachable real number.

\section{Computable and non-computable real numbers}
\label{sec:calcul}
This section recalls the concept of computable number and how the Cantor diagonal argument was used to derive a non-computable number in \cite{turing}. 

\subsection{Computable functions and computable real numbers}
A function is computable if there exists a TM which halts and prints the outputs of the function for any inputs. 
Correlatively, a real number $x$ is computable if there exists a machine which prints the $n$ first digits of $x$ for any input $n$, and  halts. 
Thus, for every computable real $x$ there exists a computable function $f$ taking an integer argument $n$ and returning the first $n$ digits of $x$.  This can be expressed by:

\begin{equation}
\forall{n}, |x-f(n)|\leq 2^{-n} \label{calc} 
\end{equation}


\subsection{Enumeration of the computable real numbers}

Each TM can be defined by a string which describes it completely (example in appendix \ref{sec:codage}), therefore the TMs are countable and can be numbered by positive integers. An example of numbering is as follows: the TMs are first ordered by the length of their description string, then the TMs with the same description lengths are sorted lexicographically. Any subset of TMs  is  countable and therefore can have its elements numbered by positive integers. 
In particular, it is possible to number all the TMs that compute real numbers. Due to the Rice theorem, a function  giving a number to each such TMs is not computable, however this function is well defined.

%
%
%

\subsection{An example of non-computable real number}
\label{sec:noncalc}
Let $\mathcal{C}_b$ be the set of TMs computing real numbers belonging to $[0,1]$ and printing their digits in base $b$, let $\alpha_k$ be the real number computed by the $k$-th  machine of $\mathcal{C}_b$  and $\phi_k(n)$ be the $n$-th digit of the number $\alpha_k$. 
Consider the real number $\beta$ such that its $n$-th digit equals $\phi_n(n)+1$ modulo $b$. For any $n$, $\beta$ differs from $\alpha_n$ by its $n$-th digit. Thus $\beta$ does not belongs to $\mathcal{C}_b$ and therefore is computable by no TM.    


Other examples of non-computable numbers are known: the Chaitin's constant $\Omega$ \cite{chaitin}; the real number such that its $n$-th digits equals 1 if a given universal TM halts for input $n$, and 0 otherwise (see\cite{janvresse}); the real number whose digits express the solutions of the busy beaver problem.

%
%
%
%
%
%
%

%


\section{Approachable real numbers}
\label{sec:app}
We say that a real number is approachable if it is the limit of a computable function, else it is not approachable.
Denoting $x$ an approachable real number and $f:\mathbb{N}\rightarrow\mathbb{R}$ a computable function approaching $x$ when argument $n\rightarrow \infty$, we have:

%
%
%
%

\begin{equation}
\forall m, \exists k\geq m, \forall n\geq k,  |x-f(n)|\leq 2^{-m} \label{app}
\end{equation}

Eq. \eqref{calc}$ \Rightarrow  $Eq. \eqref{app} 
thus any computable number is approachable. 
If $x$ is computable, $|x-f(n)|$ is computable and bounded above by $2^{-n}$. Otherwise $|x-f(n)|$ and  $x$ are not computable and  it is not possible to compute $k$ versus $m$,  nevertheless $f(n)$ is computable and approaches $x$.

We conclude that a computable real is always approachable, but the reverse is not necessarily true.

\begin{remark}If $f$ is not monotonic, $|x-f(n)|\leq 2^{-m}$ for some  $n<k$ in Eq. \ref{app}. 
\end{remark}

Formula \eqref{app} states that a real number $x$ is approachable if there exists a computable function $f$ which, for any $m$, returns $m$ digits of $x$ for any input $n$ greater than a number $k$ depending on $m$.

This can be expressed in terms of Turing machines as follows:

\begin{definition}
\label{def:app}
A TM approaches a real $x\in[0, 1]$ if:
\begin{enumerate}
	\item For any input value $n$ on the tape at initial state, the machine halts.
	\item For any $m$, there exits $k\geq m$ such that for any input $n\geq k$, a sequence of at least $m$ digits is printed on the tape when the machine halts, and the $m$ first digits of the sequence are the $m$ first digits of $x$. 
\end{enumerate}
\end{definition}

\section{An approachable number which is not computable}
\label{sec:appnoncalc}
We call ``program'' a 2-tuple defined by a TM and its input at initial state. The set of programs is countable and they can be numbered (by a computable function). Let be  the real number $h\in[0, 1]$ such that its $n$-th digit equals 1 if the $n$-th program halts, and  0 otherwise (as in \cite{janvresse}).
Since the halting problem is undecidable\cite{turing}, it is not possible to compute $h$.

However, $h$ is approachable because there exists a computable function approaching $h$.

\begin{proof}
Consider the function $H$ taking an integer argument $n$ and returning a sequence of $n$ bits such that for any $i\in [0,n[$ the $i$-th bit equals 1 if the $i$-th program halts before $n$ steps, and 0 otherwise.
The function $H$ is computable because it halts after $n^2$ steps used to run the $n$ programs during $n$ steps.
For any bit $i$ of the computed sequence: 
\begin{itemize}
	\item If the $i$-th program never halts, for any $n$, $H$ outputs the value 0 for this bit.
	\item If the $i$-th program halts at the $k$-th step, for any $n\geq k$,  $H$ outputs the value 1 for this bit.  
\end{itemize}
Therefore, for any $m$, there exists $k'$, such that for any $n\geq k'$, $H$ outputs (at least) the $m$ first bits of $h$.  
Thus $h$ is the limit of $H$ and is approachable.
\end{proof}


\begin{remark} Here $h$ is the limit of a monotonically increasing computable function.
\end{remark}

Another example of an approachable but non-computable number is the Chaitin's constant\cite{chaitin} as explained in \cite{delahaye}.
      
\section{A non-approachable number}
\label{sec:nonapp}
Consider the set $\mathcal{A}_b$ of TM approaching real numbers between 0 et 1 (definition \ref{def:app}) by printing their digits in base $b$. The set of TM and its subsets are countable, thus it is possible to number all the machines of $\mathcal{A}_b$.
We denote $\epsilon_k$ the number approached by the $k$-th machine of $\mathcal{A}_b$ and $\psi_k(n)$ the $n$-th digit of  $\epsilon_k$. Consider the real number $\gamma\in[0,1]$ such that its $n$-th digit equals $\psi_n(n)+1$ modulo $b$. For any $n$, $\gamma$ differs from $\epsilon_n$ by its $n$-th digit, so $\gamma$ does not belongs to $\mathcal{A}_b$ and $\gamma$ is not approachable.

The manner to define a non-approachable number is similar to that used in §\ref{sec:noncalc} to define a non-computable number. The difference  is that it is defined by diagonalization starting from the set of approachable numbers, rather than from the set of computable numbers.

\begin{remark} \label{rem:gamma} The number $\gamma'\in[0,1]$ such that its $n$-th digit equals $\psi_n(n)$ for any $n$ is not approachable, because else it could be used to approach $\gamma'$ and this is impossible.
\end{remark}

\section{Conclusion}
The value of $\gamma$ (or $\gamma'$) defined previously  depends on the way the TM are numbered. Appendix \ref{sec:explicite} gives an encoding and numbering of the TM which enable to define a particular value of $\gamma'$. Such a number is much more uncomputable then the Chaitin's constant $\Omega$ \cite{chaitin}. Other  well known uncomputable numbers are perhaps non-approachable (for instance, is the number $\beta$ defined in §\ref{sec:noncalc} approachable ?). Here we could display a number for which it is sure that it is not approachable.


\appendix 
\section{Definition of a non-approachable number}
We give here a description of a particular non-approachable real number. The section §\ref{sec:codage} is rather technical and deals with the encoding of a TM. 
This is then used in §\ref{sec:num} to define the value of a non-approachable real number.
\label{sec:explicite}
\subsection{Encoding of a TM}
\label{sec:codage}

Formally, a TM can be defined by a 5-tuple $M=(Q,\Gamma,\Sigma,\delta,F)$ where:
$Q$ is a finite set of states containing at least an initial state denoted $q_0$, 
$\Gamma$ is the alphabet of the tape containing at least the blank symbol $b$,
$\Sigma\subseteq\Gamma\backslash\{b\}$ is the alphabet of the inputs being on the tape to the initial state,
$\delta:Q\times\Gamma\rightarrow Q\times\Gamma\times\{L,R\}$ is the transition function where $L$ and $R$ represent respectively the left and right shift of the tape, 
$F\subseteq Q\backslash\{q_0\}$ is the finite set of final states.

Let be $n$ the number of states in $Q$, the states are numbered from 0 to $n-1$. In addition, it is assumed that the state number 0 always corresponds to the initial state $q_0$.\\
Let be  $m$ and $k$ respectively the number of symbols of the alphabets $\Gamma$ and $\Sigma$. The symbols of $\Gamma$ are numbered from  0 to $m-1$. It is assumed that the blanc symbol $b$ has number $m-1$, and the symbols of $\Sigma$ are numbered from 0 to $k-1$,  with $k\leq m-1$.  

It is assumed that to be syntactically correct machine $M$ must always contains two separate initial and final states, thus $n\geq 2$, and that the transition function $\delta$ defines one and only one transition for any configuration of $(Q\backslash F)\times \Gamma$. No further hypothesis are made about $\delta$. A minimalist machine example is: $M=(\{q_0,q_1\},\{b\},\emptyset,\{(q_0,b)\mapsto(q_0,b,L)\},\{q_1\})$. 

We choose to encode each machine with the alphabet $A=\{0, 1, (, ), `,$'$\}$ containing the 5 symbols `0', `1', `(', `)', `,'.\\
The coding of a machine begins with the symbol `$($' and ends with `$)$'. Between these two symbols are five parts separated by commas `$,$'. The first three parts are respectively the binary encodings <n>, <k>, <m> of the numbers  $n$, $k$ et $m$.\\
The fourth part encodes the transition function $\delta$. The encoding of $\delta$ begins with `$($' and ends with `$)$', the transitions are separated by `,'. Each transition $(p,a) \mapsto (q,b,x)$ of $\delta$ is encoded by the sequence of symbols (<p>,<a>,<q>,<b>,<x>) where: 
\begin{itemize}
	\item <x> is 0 if $x = L$ and is 1 if $x = R$,
	\item <p> and <q> are the binary numbers of the states $p$ and $q$ of $Q$,
	\item <a> and <b> are the binary numbers of the symbols $p$ and $q$ of  $\Gamma$.
\end{itemize}
Consider the binary number $t$ resulting from the concatenation <p>.<a>. <q>.<b>.<x>, it is assumed that the transitions of $\delta$ are ordered by increasing value of their number $t$.\\  
The last part encodes the final states of $F$. The encoding of $F$ begins with `$($' and ends with `$)$', and gives the binary numbers <f> for each state of $F$, separated by commas `$,$'. It is assumed that the states are ordered by increasing value of their number. \\
Thereby the encoding of $M$ takes the general form:\\
(<n>, <k>, <m>, ((<p1>,<a1>, <q1>,<b1>,<x1>),\\ 
(<p2>,<a2>, <q2>,<b2>,<x2>),\ldots), (<f1>, <f2>,\ldots))

A sequence of integer $N_M$ is associated to a machine $M$ by substituting the symbols 0, 1, (, ), `,' of $A$ respectively by the integers $0, 1, 2, 3, 4$. Thus, each machine $M$ whose encoding is syntactically correct is defined uniquely by a sequence of integers starting (and ending) by 2.
For instance, the minimalist machine $M=(\{q_0,q_1\},\{b\},\emptyset,\{(q_0,b)\mapsto(q_0,b,L)\},\{q_1\})$ encoded by (10,1,0,((0,0, 0,0,0)),(1)) has number $N_M=2104140422040404040224212$.

\subsection{Numbering of the approachable numbers and example of a non-approachable real number}
\label{sec:num}
Consider the set of machines $M$ approaching real numbers (définition \ref{def:app}) printed in base 10 by taking  the convention that the 10 first symbols of their alphabet $\Gamma$ correspond respectively to the numerals 0, 1, 2, 3, 4, 5, 6, 7, 8, 9 used to write numbers in base 10 (so the alphabet $\Gamma$ of these machines contains at least 11 symbols because of the blanc symbol $b$). We denote $\mathcal{A}_{10}$ this set. To any machine $M$ of $\mathcal{A}_{10}$ corresponds a sequence of integer $N_M$. This sequence is interpreted as a number (beginning with  2) written in base 5. The numbers $N_M$ are used to order the  machines of $\mathcal{A}_{10}$ by increasing number. Let $\epsilon_k$ denote the real number approached by the $k$-th machine of $\mathcal{A}_{10}$ and $\psi_k(n)$ the $n$-th digit of $\epsilon_k$ written in base 10. The number such that, for any $n$, its $n$-th digit is $\psi_n(n)$ is defined in base 10 but is not approachable (see remark \ref{rem:gamma}).

\end{document}